\documentclass[11pt]{article}
\usepackage{amssymb,amsmath,amsfonts}
\usepackage{graphicx}
\usepackage{graphics}
\usepackage{eepic,epsfig}
\usepackage{dcolumn}

\begin{document}

\title{Electromagnetic field and radiation for a charge moving along a
helical trajectory inside a waveguide with dielectric filling}
\author{A. S. Kotanjyan, A. A. Saharian \thanks{%
Email address: saharian@ictp.it} \thinspace\  \\
\textit{Institute of Applied Problems in Physics, 375014 Yerevan, Armenia}}
\date{\today}
\maketitle

\begin{abstract}
We investigate the electromagnetic field generated by a point charge moving
along a helical trajectory inside a circular waveguide with conducting walls
filled by homogeneous dielectric. The parts corresponding to the radiation
field are separated and the formulae for the radiation intensity are derived
for both TE and TM waves. It is shown that the main part of the radiated
quanta is emitted in the form of the TE waves. Various limiting cases are
considered. The results of the numerical calculations show that the
insertion of the waveguide provides an additional mechanism for tuning the
characteristics of the emitted radiation by choosing the parameters of the
waveguide and filling medium.
\end{abstract}

\bigskip

PACS number(s): 41.60.Ap, 41.60.Bq

\bigskip

\section{Introduction}

\label{sec:oscint}

A charged particle confined to the helical orbit is a source of high
intensity electromagnetic radiation over a broad range of wavelengths with a
number of remarkable properties such as high collimation and high degree of
polarization (see, for instance Refs. \cite{Soko86,Bord99,Hofm04}). These
properties have resulted in extensive applications of this radiation in a
wide variety of experiments and in many disciplines. In particular, the
helical motion of an electron beam is employed in helical undulators to
produce circularly polarized radiation in narrow angular cone in the forward
direction \cite{Alfe74,Kinc77,Luch90,Nikitin}. In the proposal of Ref. \cite%
{Bala79} the helical undulator radiation was used to generate a polarized
positron beam. Synchrotron radiation from relativistic electrons spiralling
in magnetic fields is the main mechanism to explain the emissions of many
objects in radio astronomy (see \cite{Ryb79} and references therein). Most
of the works on the radiation from the helical trajectory refer to radiation
in free space. It is well known that the presence of medium can essentially
change the characteristics of the electromagnetic processes and gives rise
to new types of phenomena such as the Cherenkov, transition, and diffraction
radiations. In particular, the operation of a number of devices assigned to
production of electromagnetic radiation is based on the interaction of
charged particles with materials (see, for example, \cite{Rull98}).

The synchrotron radiation from a charged particle circulating in a
homogeneous medium was considered in Ref. \cite{Tsytovich}. In this paper it
was shown that the interference between the synchrotron and Cherenkov
radiations leads to interesting effects. New interesting features arise in
inhomogeneous media. In particular, the interfaces of media can be used to
control the radiation flow emitted by various systems. In a series of papers
started in Refs. \cite{Grigoryan1995,Grig95b}, we have considered the
simplest geometries of boundaries between two dielectrics with spherical and
cylindrical symmetries. The synchrotron radiation from a charge rotating
around a dielectric ball enclosed by a homogeneous medium is investigated in
Refs. \cite{Grig95b,Grigoryan1998}. It was shown that if for the material of
the ball and the particle velocity the Cherenkov condition is satisfied,
strong narrow peaks appear in the radiation intensity. At these peaks the
radiated energy exceeds the corresponding quantity in a homogeneous medium
by several orders of magnitude. A similar problem with the cylindrical
symmetry has been discussed in Refs. \cite%
{Grigoryan1995,Kot2000,Kot2001,KotNIMB}. In Ref. \cite{Grigoryan1995} we
have developed a recurrent scheme for constructing the Green function of the
electromagnetic field for a medium consisting of an arbitrary number of
coaxial cylindrical layers. The investigation of the radiation from a
charged particle circulating around a dielectric cylinder immersed in a
homogeneous medium, has shown that under the Cherenkov condition for the
material of the cylinder and the velocity of the particle, there are narrow
peaks in the angular distribution of the number of quanta emitted into the
exterior space. For some values of the parameters the density of the number
of quanta in these peaks exceeds the corresponding quantity for the
radiation in vacuum by several orders. The radiation by a longitudinal
charged oscillator moving with a constant drift velocity along the axis of a
dielectric cylinder immersed in a homogeneous medium is investigated in
Refs. \cite{Saha03,Saha04}. As in the case of the circular motion it was
shown that the presence of the cylinder provides a possibility for an
essential enhancement of the radiation intensity. The properties of the
radiation from a charged particle moving along a helical orbit in
homogeneous dispersive medium are investigated in \cite{Gevo84}. The
corresponding problem for the charge moving in vacuum has been widely
discussed in literature (see, e.g., Refs. \cite{Soko86,Bord99,Hofm04,Soko68}
and references given therein). The electromagnetic field and the radiation
in the case of particle following the helical path inside a dielectric
cylinder immersed into a homogeneous medium are studied in Refs. \cite%
{Saha05,Saha06}. Recently the influence of a homogeneous
transparent medium on the radiation of relativistic particles in
planar undulators is considered in Ref. \cite{Bell06}.

In the present paper, we study the electromagnetic field and the radiation
intensity for a charge moving in a helical orbit inside a circular waveguide
with dielectric filling. Note that the radiation parts of the fields in the
case of vacuum inside the waveguide are investigated in Ref. \cite{Kara77}.
The plan of the paper is as follows. In section \ref{sec:fields} we derive
expressions for the electric and magnetic fields by making use of the
corresponding formulae from \cite{Saha06} for the geometry of a dielectric
cylinder immersed into a homogeneous medium. Analytic properties of the
corresponding Fourier components are investigated. In section \ref%
{sec:Intensity} the radiation fields are separated and they are presented as
a superposition of the waveguide eigenmodes. The formulae are derived for
the radiation intensity of TE and TM waves and numerical examples are
presented. Section \ref{sec:Conc} concludes the main results of the paper.

\bigskip

\section{Electromagnetic fields inside a waveguide}

\label{sec:fields}

Consider a point charge $q$ moving along the helical trajectory of radius $%
\rho _{0}$ inside a circular waveguide with conducting walls. We will denote
by $\rho _{1}$ the radius of the waveguide and will assume that it is filled
by homogeneous dielectric with permittivity $\varepsilon _{0}$. The particle
velocities along the axis of the waveguide (drift velocity) and in the
perpendicular plane we denote by $v_{\parallel }$ and $v_{\perp }$,
respectively. In a properly chosen cylindrical coordinate system ($\rho
,\phi ,z$) the corresponding motion is described by the coordinates
\begin{equation}
\rho =\rho _{0},\quad \phi =\omega _{0}t,\quad z=v_{\parallel }t,
\label{hetagic}
\end{equation}%
where the $z$-axis coincides with the waveguide axis and $\omega
_{0}=v_{\perp }/\rho _{0}$ is the angular velocity of the charge. This type
of motion can be produced by a uniform constant magnetic field directed
along the axis of a cylinder, by a circularly polarized plane wave, or by a
spatially periodic transverse magnetic field of constant absolute value and
a direction that rotates as a function of the coordinate $z$. In the helical
undulators the last geometry is used.

In accordance with the symmetry of the problem, we present the electric and
magnetic fields in the form of the Fourier expansion%
\begin{eqnarray}
F_{l}(\mathbf{r},t) &=&\sum_{m=-\infty }^{\infty }e^{im(\phi -\omega
_{0}t)}\int_{-\infty }^{\infty }dk_{z}e^{ik_{z}(z-v_{\parallel
}t)}F_{ml}(k_{z},\rho )  \notag \\
&=&2{\mathrm{Re}}\sideset{}{'}{\sum}_{m=0}^{\infty }e^{im(\phi -\omega
_{0}t)}\int_{-\infty }^{\infty }dk_{z}e^{ik_{z}(z-v_{\parallel
}t)}F_{ml}(k_{z},\rho ),  \label{FourFields}
\end{eqnarray}%
where $l=\rho ,\phi ,z$, and in the discussion below $F=E,H$ for the
electric and magnetic fields, respectively. The expressions for the Fourier
transforms $F_{ml}(k_{z},\rho )$ are obtained from the corresponding
formulae derived in Ref. \cite{Saha06} for the geometry of a dielectric
cylinder with permittivity $\varepsilon _{0}$ immersed into a medium with
permittivity $\varepsilon _{1}$, taking the limit $\varepsilon
_{1}\rightarrow \infty $. As a result of this limiting procedure, the
Fourier transforms are written in the decomposed form
\begin{equation}
F_{ml}(k_{z},\rho )=F_{ml}^{(0)}(k_{z},\rho )+F_{ml}^{(1)}(k_{z},\rho ),
\label{Fml}
\end{equation}%
where the part $F_{ml}^{(0)}(k_{z},\rho )$ corresponds to the fields
generated by the charge in a homogeneous medium with permittivity $%
\varepsilon _{0}$ and the part $F_{ml}^{(1)}(k_{z},\rho )$ is induced by the
presence of the waveguide. In the case $\rho <\rho _{0}$ for the homogeneous
part of the magnetic field one has%
\begin{eqnarray}
H_{ml}^{(0)} &=&-\frac{qk_{z}}{2\pi i^{\sigma _{l}}}\sum_{p=\pm 1}p^{\sigma
_{l}-1}D_{m}^{(0p)}J_{m+p}(\lambda _{0}\rho ),\;l=\rho ,\phi ,  \label{Hml0}
\\
H_{mz}^{(0)} &=&-\frac{q\lambda _{0}}{2\pi }\sum_{p=\pm
1}pD_{m}^{(0p)}J_{m}(\lambda _{0}\rho ),  \label{Hmz0}
\end{eqnarray}%
with the coefficients%
\begin{equation}
D_{m}^{(0p)}=\frac{\pi }{2ic}\left[ v_{\perp }H_{m+p}(\lambda _{0}\rho
_{0})-v_{\parallel }\frac{\lambda _{0}}{k_{z}}H_{m}(\lambda _{0}\rho _{0})%
\right] .  \label{Dm0p}
\end{equation}%
In these expressions $\sigma _{\rho }=1$, $\sigma _{\phi }=2$, $J_{m}(x)$ is
the Bessel function, $H_{m}(x)=H_{m}^{(1)}(x)$ is the Hankel function of the
first kind, and
\begin{equation}
\lambda _{0}^{2}=\frac{\omega _{m}^{2}(k_{z})}{c^{2}}\varepsilon
_{0}-k_{z}^{2},\quad \omega _{m}(k_{z})=m\omega _{0}+k_{z}v_{\parallel }.
\label{lambdaj}
\end{equation}%
The corresponding expressions for $\rho <\rho _{0}$ are obtained from (\ref%
{Hml0}), (\ref{Hmz0}) by the replacements $J\rightleftarrows H$. The part $%
H_{ml}^{(1)}(k_{z},\rho )$ induced by the presence of the waveguide is given
by the formulae
\begin{eqnarray}
H_{ml}^{(1)} &=&-\frac{qk_{z}}{2\pi i^{\sigma _{l}}}\sum_{p=\pm 1}p^{\sigma
_{l}-1}D_{m}^{(p)}J_{m+p}(\lambda _{0}\rho ),\;l=\rho ,\phi ,  \label{Hml1w}
\\
H_{mz}^{(1)} &=&-\frac{q\lambda _{0}}{2\pi }\sum_{p=\pm
1}pD_{m}^{(p)}J_{m}(\lambda _{0}\rho ),  \label{Hmz1w}
\end{eqnarray}%
where we have introduced the notation%
\begin{eqnarray}
D_{m}^{(p)} &=&\frac{\pi }{2ic}\left[ v_{\parallel }\frac{\lambda _{0}}{k_{z}%
}\frac{H_{m}(\lambda _{0}\rho _{1})}{J_{m}(\lambda _{0}\rho _{1})}%
J_{m}(\lambda _{0}\rho _{0})-v_{\perp }J_{m+p}(\lambda _{0}\rho _{0})\frac{%
H_{m+p}(\lambda _{0}\rho _{1})}{J_{m+p}(\lambda _{0}\rho _{1})}\right.
\notag \\
&&\left. -\frac{iv_{\perp }p}{\pi \rho _{1}\lambda _{0}}\frac{%
J_{m-p}(\lambda _{0}\rho _{1})}{J_{m}(\lambda _{0}\rho _{1})J_{m}^{\prime
}(\lambda _{0}\rho _{1})}\sum_{l=\pm 1}l\frac{J_{m+l}(\lambda _{0}\rho _{0})%
}{J_{m+l}(\lambda _{0}\rho _{1})}\right] ,  \label{Dmpw}
\end{eqnarray}%
and the prime stands for the derivative with respect to the argument of the
function.

The part in the electric field due to the presence of the waveguide is given
by the formulae
\begin{eqnarray}
E_{ml}^{(1)} &=&\frac{qci^{1-\sigma _{l}}}{4\pi \omega
_{m}(k_{z})\varepsilon _{0}}\sum_{p=\pm 1}p^{\sigma _{l}}J_{m+p}(\lambda
_{0}\rho )\left[ \left( \frac{\omega _{m}^{2}(k_{z})\varepsilon _{0}}{c^{2}}%
+k_{z}^{2}\right) D_{m}^{(p)}-\lambda _{0}^{2}D_{m}^{(-p)}\right] ,
\label{electricb} \\
E_{mz}^{(1)} &=&\frac{qic\lambda _{0}k_{z}}{2\pi \omega
_{m}(k_{z})\varepsilon _{0}}\sum_{p=\pm 1}D_{m}^{(p)}J_{m}(\lambda _{0}\rho
),  \label{electricc}
\end{eqnarray}%
where $l=\rho ,\phi $. The corresponding formulae for the part $E_{ml}^{(0)}$
are obtained from (\ref{electricb}), (\ref{electricc}) by the replacement $%
D_{m}^{(p)}\rightarrow D_{m}^{(0p)}$. Note that in the limit $\rho
_{0}\rightarrow \rho _{1}$ one has $D_{m}^{(p)}\rightarrow -D_{m}^{(0p)}$
and the fields vanish. We could expect this result, as when the charge is on
the surface of the waveguide the charge and its image compensate each other.

The formulae given above describe the total electromagnetic field inside the
waveguide. To separate the parts corresponding to the radiation we need the
analytic properties of the Fourier components as functions on $k_{z}$. From
formula (\ref{Dmpw}) it follows that the function $D_{m}^{(p)}$ has singular
points corresponding to the zeros of the function $J_{m}(\lambda _{0}\rho
_{1})$ and its derivative. Note that in the second and third summands on the
right of formula (\ref{Dmpw}) the singularities at the zeros of the
functions $J_{m\pm 1}(\lambda _{0}\rho _{1})$ cancel out and the function $%
D_{m}^{(p)}$ is analytic at these points. We denote by $j_{m,n}^{(\sigma )}$%
, $n=1,2,\ldots $, the $n$th positive zero of the Bessel function ($\sigma
=0 $) and its derivative ($\sigma =1$):
\begin{equation}
J_{m}^{(\sigma )}(\lambda _{0}\rho _{1})=J_{m}^{(\sigma )}(j_{m,n}^{(\sigma
)})=0,  \label{Besselzeros}
\end{equation}%
where $J_{m}^{(\sigma )}(x)=d^{\sigma }J_{m}(x)/dx^{\sigma }$. These zeros
describe the eigenmodes for the cylindrical waveguide and are known as TM
modes for the case $\sigma =0$ and as TE modes for the case $\sigma =1$ \cite%
{Jackson}. The corresponding modes for the projection of the wave vector on
the cylinder axis are determined from the relation $\lambda _{0}\rho
_{1}=j_{m,n}^{(\sigma )}$ by taking into account expressions (\ref{lambdaj}):%
\begin{equation}
\left( \beta _{\parallel }^{2}-1\right) k_{z}^{2}+2m\frac{\omega _{0}}{c}%
\sqrt{\varepsilon _{0}}\beta _{\parallel }k_{z}+\left( m^{2}\frac{\omega
_{0}^{2}\varepsilon _{0}}{c^{2}}-\frac{j_{m,n}^{(\sigma )2}}{\rho _{1}^{2}}%
\right) =0,\;\beta _{\parallel }=\frac{v_{\parallel }}{c}\sqrt{\varepsilon
_{0}}.  \label{eqforkz}
\end{equation}%
This equation has real solutions under the condition%
\begin{equation}
b_{m,n}^{(\sigma )2}\left( 1-\beta _{\parallel }^{2}\right) \leqslant
1,\;b_{m,n}^{(\sigma )}\equiv \frac{cj_{m,n}^{(\sigma )}}{m\omega _{0}\rho
_{1}\sqrt{\varepsilon _{0}}},  \label{solcond}
\end{equation}%
and these solutions have the form%
\begin{equation}
k_{z}=k_{m,n}^{(\sigma ,\pm )}=\frac{m\omega _{0}\sqrt{\varepsilon _{0}}}{%
c\left( 1-\beta _{\parallel }^{2}\right) }\left[ \beta _{\parallel }\pm
\sqrt{1+b_{m,n}^{(\sigma )2}\left( \beta _{\parallel }^{2}-1\right) }\right]
.  \label{kmnpm}
\end{equation}%
The values $k_{m,n}^{(\sigma ,\pm )}$ correspond to the simple poles of the
functions $E_{ml}(k_{z},\rho )$ and $H_{ml}(k_{z},\rho )$. Note that if the
Cherenkov condition for the velocity of the particle along the axis of the
waveguide is satisfied, $\beta _{\parallel }>1$, then inequality (\ref%
{solcond}) is valid for all values $n=1,2,\ldots $. In the case $\beta
_{\parallel }<1$, condition (\ref{solcond}) determines the maximal value for
$n$, which we will denote by $n_{\max }^{(\sigma )}$:%
\begin{equation}
j_{m,n_{\max }^{(\sigma )}}^{(\sigma )}<\frac{m\omega _{0}\rho _{1}\sqrt{%
\varepsilon _{0}}}{c\sqrt{1-\beta _{\parallel }^{2}}}<j_{m,n_{\max
}^{(\sigma )}+1}^{(\sigma )}.  \label{defnmax}
\end{equation}%
The value of the function $\omega _{m}(k_{z})$ at the points $%
k_{m,n}^{(\sigma ,\pm )}$ is equal to%
\begin{equation}
\omega _{m,n}^{(\sigma ,\pm )}=\frac{m\omega _{0}}{1-\beta _{\parallel }^{2}}%
\left[ 1\pm \beta _{\parallel }\sqrt{1+b_{m,n}^{(\sigma )2}\left( \beta
_{\parallel }^{2}-1\right) }\right] .  \label{ommnsigw}
\end{equation}%
For $m=0$, equation (\ref{eqforkz}) has real solutions only when $\beta
_{\parallel }>1$ and
\begin{equation}
k_{z}=k_{0,n}^{(\sigma ,\pm )}=\mp \frac{j_{0,n}^{(\sigma )}}{\rho _{1}\sqrt{%
\beta _{\parallel }^{2}-1}},\;\omega _{0,n}^{(\sigma ,\pm )}=v_{\parallel
}k_{0,n}^{(\sigma ,\pm )}.  \label{kzm0w}
\end{equation}

\begin{figure}[tbph]
\begin{center}
\epsfig{figure=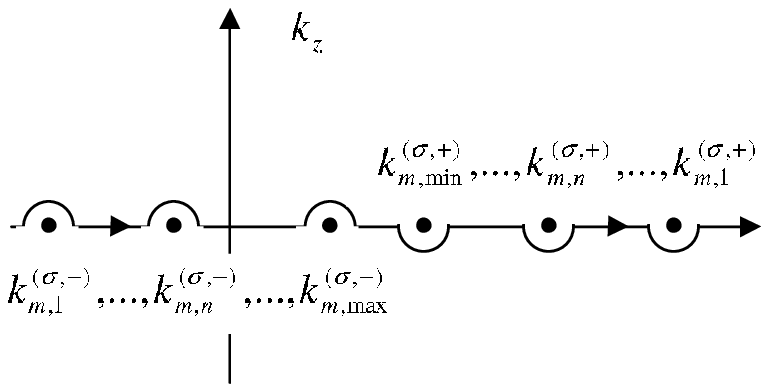,width=7.5cm,height=4cm}
\end{center}
\caption{Integration contour in the $k_z$ plane for the case $\protect\beta %
_{\parallel }<1$.}
\label{fig1}
\end{figure}

\begin{figure}[tbph]
\begin{center}
\epsfig{figure=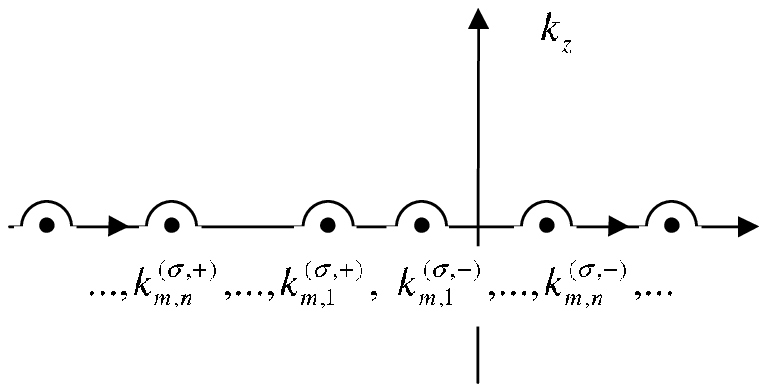,width=7.5cm,height=4cm}
\end{center}
\caption{Integration contour in the $k_z$ plane for the case $\protect\beta %
_{\parallel }>1$.}
\label{fig2}
\end{figure}

For real values $\varepsilon _{0}$ and under the condition (\ref{solcond})
the poles are situated on the real axis of the complex plane $k_{z}$. In
formula (\ref{FourFields}) it is necessary to give the way by which these
poles should be circled. For this we note that in physical situations the
dielectric permittivity is a complex quantity, $\varepsilon _{0}=\varepsilon
_{0}^{\prime }+i\varepsilon _{0}^{\prime \prime }$, and the imaginary part $%
\varepsilon _{0}^{\prime \prime }$ determines the absorbtion in the medium.
Under the condition $|\varepsilon _{0}^{\prime \prime }|\ll \varepsilon
_{0}^{\prime }$, for the imaginary part of $k_{m,n}^{(\sigma ,\pm )}$ from (%
\ref{kmnpm}) one has%
\begin{equation}
{\mathrm{Im\,}}k_{m,n}^{(\sigma ,\pm )}=\pm C_{1}\varepsilon _{0}^{\prime
\prime }(\omega _{m}),\;C_{1}>0,  \label{Imkmnw}
\end{equation}%
where $\varepsilon _{0}^{\prime \prime }(\omega _{m})\gtrless 0$ for $\omega
_{m}\gtrless 0$. It can be also seen that $(b_{m,n}^{(\sigma )}-1){\mathrm{%
Re\,}}k_{m,n}^{(\sigma ,-)}>0$. If the Cherenkov condition is not satisfied,
$\beta _{\parallel }=v_{\parallel }\sqrt{\varepsilon _{0}^{\prime }}/c<1$,
from the formulae given above it follows that
\begin{equation}
{\mathrm{Re\,}}k_{m,n}^{(\sigma ,+)}>0,\;\omega _{m}({\mathrm{Re\,}}%
k_{m,n}^{(\sigma ,\pm )})>0,\;\pm {\mathrm{Im\,}}k_{m,n}^{(\sigma ,\pm )}>0.
\label{case1}
\end{equation}%
In this case the poles $k_{m,n}^{(\sigma ,+)}$ ($k_{m,n}^{(\sigma ,-)}$) are
situated in the upper (lower) half of the complex plane $k_{z}$. In the
limit $\varepsilon _{0}^{\prime \prime }\rightarrow 0$, deforming the
integration contour we obtain the rule for avoiding the poles plotted in
figure \ref{fig1}. In the case $\beta _{\parallel }>1$ the corresponding
inequalities have the form%
\begin{equation}
{\mathrm{Re\,}}k_{m,n}^{(\sigma ,+)}<0,\;\mp \omega _{m}({\mathrm{Re\,}}%
k_{m,n}^{(\sigma ,\pm )})<0,\;{\mathrm{Im\,}}k_{m,n}^{(\sigma ,\pm )}<0.
\label{case2}
\end{equation}%
In the way similar to that for the previous case, deforming the contour for
the integration over $k_{z}$, we obtain the avoidance rule for the poles
given in figure \ref{fig2}.


\section{Radiation intensity in the waveguide}

\label{sec:Intensity}

\subsection{Radiation fields}

In this section we consider the radiation field propagating inside a
cylindrical waveguide with perfectly conducting walls for large distances
from the charge. First of all let us show that the part corresponding to the
first term on the right of formula (\ref{Fml}) does not contribute to the
radiation field. This directly follows from the estimate of the integral
over $k_{z}$ on the base of the stationary phase method. As in the integral
over $k_{z}$ the phase $k_{z}z$ has no stationary points, for large values $%
|z|$ the integral vanishes faster than any degree of $1/|z|$, under the
condition that the preexponential function belongs to the class $C^{\infty
}(R)$. It follows from here that the radiation field is determined by the
singularity points of the preexponential function. As it has been mentioned
before, for the integral over $k_{z}$ the only singular points are the poles
of the function $D_{m}^{(p)}$ at the points $k_{z}=k_{m,n}^{(\sigma ,\pm )}$%
, determined by relations (\ref{kmnpm}). To find the corresponding
contribution, we note that the integration contour over $k_{z}$ has the form
depicted in figure \ref{fig1} for $\beta _{\parallel }<1$ and the form
depicted in figure \ref{fig2} for the case $\beta _{\parallel }>1$. At large
distances from the charge the integration contour can be closed by large
semicircle in the upper (lower) half-plane for $z>v_{\parallel }t$ ($%
z<v_{\parallel }t$). This choice is caused by the fact that the integrand
exponentially vanishes in the upper (lower) half-plane. As a result when the
radius of the large semicircle goes to infinity the corresponding integral
vanishes. Hence, for large values $|z|$ the integral over $k_{z}$, by
residue theorem, is equal to the sum of residues inside the contour
multiplied by $2\pi i\mathrm{sgn\,}(z-v_{\parallel }t)$. For $\varepsilon
_{0}^{\prime \prime }\rightarrow 0$ and large $n$, when $b_{m,n}^{(\sigma
)2}>\left( 1-\beta _{\parallel }^{2}\right) $ $^{-1}$, the poles $%
k_{m,n}^{(\sigma ,\pm )}$ have a finite imaginary part and the corresponding
contribution exponentially vanishes in the limit $z\rightarrow \infty $. As
a result these poles do not contribute to the radiation field.

As before we consider two cases. When $\beta _{\parallel }<1$ we close the
integration contour in figure \ref{fig1} by the semicircle of large radius
in the upper half-plane for $z>v_{\parallel }t$ and in the lower half-plane
for $z<v_{\parallel }t$. As a result for the radiation field one finds
\begin{equation}
F_{l}(\mathbf{r},t)=\alpha {\mathrm{Re}}\left[ \sum_{\sigma =0,1}%
\sideset{}{'}{\sum}_{m=0}^{\infty }\sum_{n=1}^{n_{\max }^{(\sigma
)}}F_{ml}^{(\sigma ,\alpha )}(\mathbf{r},t)\right] ,  \label{radfieldw}
\end{equation}%
where $\alpha =+$ $\ $($\alpha =-$) corresponds to the case $z>v_{\parallel
}t$ ($z<v_{\parallel }t$) and we use the notation%
\begin{equation}
F_{ml}^{(\sigma ,\alpha )}(\mathbf{r},t)=4\pi i\underset{k_{z}=k_{m,n}^{(%
\sigma ,\alpha )}}{\mathrm{Res}}F_{ml}(k_{z},\rho )e^{i(m\varphi
+k_{z}z-\omega _{m}t)}.  \label{Resnot}
\end{equation}%
Each term in the sum on the right of formula (\ref{radfieldw}) describes
waves with the frequency $\omega _{m,n}^{(\sigma ,\pm )}$ propagating along
the positive direction of the axis $z$ for $\alpha =+$ and for $\alpha =-$, $%
1<b_{m,n}^{(\sigma )}<1/\sqrt{1-\beta _{\parallel }^{2}}$, and waves
propagating along the negative direction of the axis $z$ for $\alpha =-$, $%
b_{m,n}^{(\sigma )}<1$. If the Cerenkov condition is satisfied, $\beta
_{\parallel }>1$, then closing the integration contour in figure \ref{fig2}
by large semicircle in the upper or lower half-plane in dependence of the
sign for $z-v_{\parallel }t$, for the vector potential of the radiation
field one finds%
\begin{equation}
F_{l}(\mathbf{r},t)=-\theta (v_{\parallel }t-z){\mathrm{Re}}\left[
\sum_{\alpha =\pm }\sum_{\sigma =0,1}\sideset{}{'}{\sum}_{m=0}^{\infty
}\sum_{n=1}^{n_{\max }^{(\sigma )}}F_{ml}^{(\sigma ,\alpha )}(\mathbf{r},t)%
\right] ,  \label{radfieldw2}
\end{equation}%
where $\theta (x)$ is the Heaviside unit step function. Separate terms in
the sum in formula (\ref{radfieldw2}) describe waves with the frequency $%
|\omega _{m,n}^{(\sigma ,\pm )}|$ propagating along the positive direction
of the axis $z$ for $\alpha =+$ and for $\alpha =-$, $b_{m,n}^{(\sigma )}>1$%
, and along the negative direction of the axis $z$ for $\alpha =-$, $%
b_{m,n}^{(\sigma )}<1$. Note that for $b_{m,n}^{(\sigma )}>1$ we have no
waves propagating along the negative direction of the axis $z$. By taking
into account the formulae for the Fourier components of the fields and
evaluating the residues by the standard formulae of the complex analysis, we
find the following expressions for the radiation parts of the $z$-components
of the fields:%
\begin{eqnarray}
E_{mz}^{(0,\alpha )}(\mathbf{r},t) &=&\frac{4qJ_{m}(j_{m,n}^{(0)}\rho
_{0}/\rho _{1})}{\varepsilon _{0}\rho _{1}^{2}J_{m+1}^{2}(j_{m,n}^{(0)})}%
J_{m}(j_{m,n}^{(0)}\rho /\rho _{1})\exp [i(m\varphi +k_{m,n}^{(0,\alpha
)}z-\omega _{m,n}^{(0,\alpha )}t)],  \label{Ez0} \\
H_{mz}^{(1,\alpha )}(\mathbf{r},t) &=&-\alpha \frac{4iqv_{\perp
}j_{m,n}^{(1)3}J_{m}^{\prime }(j_{m,n}^{(1)}\rho _{0}/\rho _{1})}{\sqrt{%
\varepsilon _{0}}\rho _{1}^{3}(j_{m,n}^{(1)}-m^{2})J_{m}^{2}(j_{m,n}^{(1)})}%
J_{m}(j_{m,n}^{(1)}\rho /\rho _{1})  \notag \\
&&\frac{\exp [i(m\varphi +k_{m,n}^{(1,\alpha )}z-\omega _{m,n}^{(1,\alpha
)}t)]}{m\omega _{0}\sqrt{1+b_{m,n}^{(\sigma )2}\left( \beta _{\parallel
}^{2}-1\right) }},  \label{Hz1}
\end{eqnarray}%
and $H_{mz}^{(0,\alpha )}(\mathbf{r},t)=E_{mz}^{(1,\alpha )}(\mathbf{r},t)=0$%
. The transverse components are found from the formulae%
\begin{eqnarray}
\mathbf{E}_{mt}^{(0,\alpha )}(\mathbf{r},t) &=&\frac{ik_{m,n}^{(0,\alpha )}}{%
j_{m,n}^{(0)2}}\rho _{1}^{2}\mathbf{\nabla }_{t}\psi ^{(0)},\;\mathbf{H}%
_{mt}^{(0,\alpha )}(\mathbf{r},t)=\frac{\varepsilon _{0}\omega
_{m,n}^{(0,\alpha )}}{ck_{m,n}^{(0,\alpha )}}\mathbf{e}_{3}\times \mathbf{E}%
_{mt}^{(0,\alpha )}(\mathbf{r},t),  \label{Etrans} \\
\mathbf{H}_{mt}^{(1,\alpha )}(\mathbf{r},t) &=&\frac{ik_{m,n}^{(1,\alpha )}}{%
j_{m,n}^{(1)2}}\rho _{1}^{2}\mathbf{\nabla }_{t}\psi ^{(1)},\;\mathbf{E}%
_{mt}^{(1,\alpha )}(\mathbf{r},t)=-\frac{\omega _{m,n}^{(1,\alpha )}}{%
ck_{m,n}^{(1,\alpha )}}\mathbf{e}_{3}\times \mathbf{H}_{mt}^{(1,\alpha )}(%
\mathbf{r},t),  \label{Htrans}
\end{eqnarray}%
where $\psi ^{(0)}=E_{mz}^{(0,\alpha )}(\mathbf{r},t)$ and $\psi
^{(1)}=H_{mz}^{(1,\alpha )}(\mathbf{r},t)$, $\mathbf{\nabla }_{t}=(\partial
/\partial \rho ,im/\rho ,0)$, and $\mathbf{e}_{3}$ is the unit vector along
the axis of the waveguide.

\subsection{Radiation intensity}

As we have seen, inside the waveguide the radiation is presented in the form
of waves with discrete set of the values for the projection of the wave
vector on the waveguide axis, $k_{z}=k_{m,n}^{(\sigma ,\alpha )}$, $%
n=1,2,\ldots $, determined by formula (\ref{kmnpm}). Having the radiation
fields we consider the mean energy lost per unit time,%
\begin{equation}
I=-\frac{1}{T}\int_{0}^{T}dt\int \left( j_{\varphi }E_{\varphi
}+j_{z}E_{z}\right) \rho d\rho d\varphi dz,\;T=2\pi /\omega _{0}.
\label{Int1}
\end{equation}%
The radiation intensity $I$ is presented as the sum of intensities on
separate modes:%
\begin{equation}
I=\sideset{}{'}{\sum}_{m=0}^{\infty }\sum_{\sigma =0,1}I_{m}^{(\sigma
)},\;I_{m}^{(\sigma )}=\sum_{\alpha =\pm }\sum_{n=1}^{n_{\max }^{(\sigma
)}}I_{m,n}^{(\sigma ,\alpha )}.  \label{Int2}
\end{equation}%
The term with $m=0$ is present only then the condition $\beta _{\parallel
}>1 $ is satisfied and the corresponding parts have the form%
\begin{equation}
I_{0}^{(0)}=\frac{2q^{2}v_{\parallel }}{\rho _{1}^{2}}\sum_{n=1}^{n_{\max
}^{(0)}}\frac{J_{0}^{2}(j_{0,n}^{(0)}\rho _{0}/\rho _{1})}{\varepsilon
_{0}J_{1}^{2}(j_{0,n}^{(0)})},\;I_{0}^{(1)}=\frac{2q^{2}v_{\perp
}^{2}v_{\parallel }}{c^{2}\rho _{1}^{2}}\sum_{n=1}^{n_{\max }^{(1)}}\frac{%
J_{1}^{2}(j_{0,n}^{(1)}\rho _{0}/\rho _{1})}{\left( \beta _{\parallel
}^{2}-1\right) J_{0}^{2}(j_{0,n}^{(1)})}.  \label{Im0}
\end{equation}%
Each term in the sums of these formulae corresponds to the radiation with
the frequency given by the formula $v_{\parallel }j_{0,n}^{(\sigma )}/(\rho
_{1}\sqrt{\beta _{\parallel }^{2}-1})$. For the radiation intensities on
harmonics $m\neq 0$ one has the formulae%
\begin{eqnarray}
I_{m,n}^{(0,\alpha )} &=&\frac{2q^{2}c}{\varepsilon _{0}^{3/2}\rho _{1}^{2}}%
\frac{J_{m}^{2}(j_{m,n}^{(0)}\rho _{0}/\rho _{1})}{%
b_{m,n}^{(0)2}J_{m+1}^{2}(j_{m,n}^{(0)})}\frac{|\omega _{m,n}^{(0,\alpha )}|%
}{m\omega _{0}}\sqrt{1+b_{m,n}^{(0)2}\left( \beta _{\parallel }^{2}-1\right)
},  \label{ITM} \\
I_{m,n}^{(1,\alpha )} &=&\frac{2q^{2}v_{\perp }^{2}}{c\sqrt{\varepsilon _{0}}%
\rho _{1}^{2}}\frac{j_{m,n}^{(1)2}J_{m}^{^{\prime }2}(j_{m,n}^{(1)}\rho
_{0}/\rho _{1})}{\left( j_{m,n}^{(1)2}-m^{2}\right) J_{m}^{2}(j_{m,n}^{(1)})}%
\frac{|\omega _{m,n}^{(1,\alpha )}|}{m\omega _{0}\sqrt{1+b_{m,n}^{(1)2}%
\left( \beta _{\parallel }^{2}-1\right) }}.  \label{ITE}
\end{eqnarray}%
For $\beta _{\parallel }<1$ the upper limit of the summation over $n$ is
defined by relation (\ref{defnmax}). Otherwise this limit is determined by
the dispersion law for the dielectric permittivity $\varepsilon _{0}$
through the condition $v_{\parallel }\sqrt{\varepsilon _{0}}>c$. It can be
seen that for the case $\varepsilon _{0}=1$ the expression for $\sum_{\alpha
}I_{m,n}^{(\sigma ,\alpha )}$ obtained from formulae (\ref{ITM}) and (\ref%
{ITE}) coincides with the corresponding formulae in Ref. \cite{Kara77}.
Taking $\beta _{\parallel }=0$, from formulae (\ref{ITM}) and (\ref{ITE}) we
obtain the corresponding results for the radiation from a particle
circulating in the plane perpendicular to the waveguide axis \cite{Kota02}.

In accordance with (\ref{solcond}), for given $m$ and $n$ the necessary
condition for the presence of the radiation is the condition%
\begin{equation}
m\omega _{0}\rho _{1}\sqrt{\varepsilon _{0}}/c\geqslant j_{m,n}^{(\sigma )}%
\sqrt{1-\beta _{\parallel }^{2}}.  \label{radpres}
\end{equation}%
Now by taking into account the relation $j_{m,n}^{(\sigma )}\geqslant m$ for
the zeros of the Bessel functions, we conclude that under the conditions $%
\beta _{\parallel }<1$ and $\beta _{\perp }<\sqrt{1-\beta _{\parallel }^{2}}%
\rho _{0}/\rho _{1}$, with $\beta _{\perp }=v_{\perp }\sqrt{\varepsilon _{0}}%
/c$, there is no radiation inside the waveguide though the particle moves
with acceleration. If $\beta _{\parallel }<1$ and the condition%
\begin{equation}
j_{m,n}^{(\sigma )}\sqrt{1-\beta _{\parallel }^{2}}=m\beta _{\perp }\rho
_{1}/\rho _{0}  \label{peakcond}
\end{equation}
takes place then the intensity for the TE waves defined by formulae (\ref%
{ITE}) goes to infinity. However, under these conditions the absorption in
the medium (and also in the walls of the waveguide) becomes important and
the imaginary part of the dielectric permittivity should be taken into
account. Formulae (\ref{ITM}) and (\ref{ITE}) are valid \ under the condition%
\begin{equation}
\frac{\varepsilon _{0}^{\prime \prime }}{\varepsilon _{0}^{\prime }}\ll
\varepsilon _{0}^{\prime }\left[ \left( \frac{m\omega _{0}\rho _{1}}{%
j_{m,n}^{(\sigma )}c}\right) ^{2}+\frac{v_{\parallel }^{2}}{c^{2}}\right] -1,
\label{conddiv}
\end{equation}%
where $\varepsilon _{0}=\varepsilon _{0}^{\prime }+i\varepsilon _{0}^{\prime
\prime }$.

Instead of $k_{z}$ we can introduce the angular variable $\vartheta $, the
values $\vartheta _{m,n}^{(\sigma ,\alpha )}$ for which are related with the
quantities $k_{m,n}^{(\sigma ,\alpha )}$ by the formula%
\begin{equation}
k_{m,n}^{(\sigma ,\alpha )}=\frac{m\omega _{0}}{c}\frac{\sqrt{\varepsilon
_{0}}\cos \vartheta _{m,n}^{(\sigma ,\alpha )}}{1-\beta _{\parallel }\cos
\vartheta _{m,n}^{(\sigma ,\alpha )}}.  \label{teta}
\end{equation}%
The quantities $\omega _{m}(k_{z})$ and $\lambda _{0}$ are expressed via $%
\vartheta _{m,n}^{(\sigma ,\alpha )}$ by the formulae%
\begin{eqnarray}
\omega _{m}(k_{z}) &=&\frac{m\omega _{0}}{1-\beta _{\parallel }\cos
\vartheta _{m,n}^{(\sigma ,\alpha )}},  \label{omegamw} \\
\lambda _{0} &=&\frac{\omega _{m}(k_{z})}{c}\sqrt{\varepsilon _{0}}\sin
\vartheta _{m,n}^{(\sigma ,\alpha )}.  \label{lam0w}
\end{eqnarray}%
The possible values $\vartheta _{m,n}^{(\sigma ,\alpha )}$ are determined by
formulae (\ref{kmnpm}) and (\ref{teta}):%
\begin{equation}
\cos \vartheta _{m,n}^{(\sigma ,\pm )}=\frac{\beta _{\parallel
}b_{m,n}^{(\sigma )2}\pm \sqrt{1+b_{m,n}^{(\sigma )2}\left( \beta
_{\parallel }^{2}-1\right) }}{1+\beta _{\parallel }^{2}b_{m,n}^{(\sigma )2}}.
\label{costet}
\end{equation}%
Note that the singularity in the radiation intensity (\ref{ITE}) noted above
corresponds to the values of the angular variable determined by the condition%
\begin{equation}
\vartheta _{m,n}^{(\sigma ,\alpha )}=\vartheta _{\perp },\;\vartheta _{\perp
}\equiv \arccos \beta _{\parallel }.  \label{tetdiv}
\end{equation}%
In the reference frame moving along the direction of the axis $z$ with the
velocity $v_{\parallel }$ the angle corresponding to $\vartheta _{\perp }$
is equal to $\pi /2$. From (\ref{costet}) the following relations can be seen%
\begin{equation}
0\leqslant \vartheta _{m,n}^{(\sigma ,+)}\leqslant \vartheta _{0}\leqslant
\vartheta _{m,n}^{(\sigma ,-)}\leqslant \pi ,  \label{angreg}
\end{equation}%
where%
\begin{equation}
\vartheta _{0}=\left\{
\begin{array}{cc}
\vartheta _{\perp } & \mathrm{for}\;\beta _{\parallel }<1, \\
\vartheta _{c} & \mathrm{for}\;\beta _{\parallel }>1,%
\end{array}%
\right.  \label{thet0}
\end{equation}%
and $\vartheta _{c}=\arccos (1/\beta _{\parallel })$ is the Cherenkov angle
related to the drift velocity $v_{\parallel }$. Hence, for $\beta
_{\parallel }<1$ the waves with $\alpha =+$ ($\alpha =-$) are those which in
the reference frame moving with velocity $v_{\parallel }$ along the
waveguide axis, propagate along the positive (negative) direction of the
axis $z$. For $\beta _{\parallel }>1$ the waves with $\alpha =+$ ($\alpha =-$%
) propagate inside (outside) the Cherenkov cone $\vartheta =\vartheta _{c}$.
From formulae (\ref{ITM}), (\ref{ITE}) it follows that the number of
radiated quanta does not depend on $\alpha $. In particular, for $\beta
_{\parallel }<1$, the same number of quanta is radiated inside and outside
the cone $\vartheta =\vartheta _{\perp }$. For the case $\varepsilon _{0}=1$
we could expect this result from the problem symmetry, as in the reference
frame moving along the direction of the axis $z$ with the velocity $%
v_{\parallel }$ we have a symmetric situation under the reflection with
respect to the charge rotation plane.

Now let us consider the radiation intensity in the limiting case of large
values of the waveguide radius, $\rho _{1}\rightarrow \infty $. In this
limit the main contribution into the radiation intensity comes from large
values $n$ and we can use the asymptotic formula (see, for instance, \cite%
{Abramovic})
\begin{equation}
j_{m,n}^{(\sigma )}\approx \pi \left( n+\frac{m-1}{2}+\frac{(-1)^{\sigma }}{4%
}\right) .  \label{jmnas}
\end{equation}%
Replacing the summation over $n$ by the integration and introducing as a new
integration variable the angle $\vartheta $, for the the radiation intensity
one finds
\begin{equation}
I_{m}^{(\sigma )}\approx \int d\vartheta \frac{dI_{0m}^{(\sigma )}}{%
d\vartheta },  \label{Imsigma}
\end{equation}%
where%
\begin{eqnarray}
\frac{dI_{0m}^{(0)}}{d\vartheta } &=&\frac{q^{2}\omega _{0}^{2}m^{2}}{c\sqrt{%
\varepsilon _{0}}\sin \vartheta }\frac{(\cos \vartheta -\beta _{\parallel
})^{2}}{|1-\beta _{\parallel }\cos \vartheta |^{3}}J_{m}^{2}\left( \frac{%
m\beta _{\perp }\sin \vartheta }{1-\beta _{\parallel }\cos \vartheta }%
\right) ,  \label{ImhamTM} \\
\frac{dI_{0m}^{(1)}}{d\vartheta } &=&\frac{q^{2}\omega _{0}^{2}m^{2}}{c\sqrt{%
\varepsilon _{0}}}\frac{\beta _{\perp }^{2}\sin \vartheta }{|1-\beta
_{\parallel }\cos \vartheta |^{3}}J_{m}^{\prime 2}\left( \frac{m\beta
_{\perp }\sin \vartheta }{1-\beta _{\parallel }\cos \vartheta }\right) .
\label{ImhamTE}
\end{eqnarray}%
In this limit the frequency for the radiation along given direction $%
\vartheta $ is determined by the expression $m\omega _{0}/|1-\beta
_{\parallel }\cos \vartheta |$. The expressions (\ref{ImhamTM}) and (\ref%
{ImhamTE}) coincide with the corresponding formulae for the radiation in a
homogeneous medium with dielectric permittivity $\varepsilon _{0}$. Note
that in the discussed limit we have the transition $\vartheta
_{m,n}^{(\sigma ,\pm )}\rightarrow \vartheta $ and the upper (lower) sign
corresponds to the angular region $0\leqslant \vartheta \leqslant \vartheta
_{0}$ ($\vartheta _{0}\leqslant \vartheta \leqslant \pi $), where $\vartheta
_{0}$ is defined by formula (\ref{thet0}).

As an additional check for formulae (\ref{ITM}), (\ref{ITE}) we can consider
the special case $\omega _{0}=0$ for a fixed value $\rho _{0}$. This
corresponds to a charge moving with constant velocity $v_{\parallel }$ on a
straight line $\rho =\rho _{0}$ parallel to the waveguide axis. In this
limit $b_{m,n}^{(\sigma )}\rightarrow \infty $ and from condition (\ref%
{solcond}) it follows that the radiation is present only under the Cherenkov
condition $\beta _{\parallel }>1$. Taking the limit $\omega _{0}\rightarrow
0 $, from formulae (\ref{ITM}), (\ref{ITE}) we see that
\begin{equation}
I_{m,n}^{(0,\alpha )}|_{\omega _{0}=0}=\frac{2q^{2}v_{\parallel }}{%
\varepsilon _{0}\rho _{1}^{2}}\frac{J_{m}^{2}(j_{m,n}^{(0)}\rho _{0}/\rho
_{1})}{J_{m+1}^{2}(j_{m,n}^{(0)})},\;I_{m,n}^{(1,\alpha )}\rightarrow 0.
\label{Omega0}
\end{equation}%
Hence, in the limit under consideration the TM waves are radiated only. The
corresponding frequency is given by the expression $v_{\parallel
}j_{m,n}^{(0)}/(\rho _{1}\sqrt{\beta _{\parallel }^{2}-1})$. In the limit $%
\rho _{1}\rightarrow \infty $, by using asymptotic formula (\ref{jmnas}) for
the zeros of the Bessel function, replacing the summation over $n$ by the
integration, and using the formula $\sideset{}{'}{\sum}_{m=0}^{\infty
}J_{m}^{2}(x)=1/2$, we can see that from (\ref{Omega0}) the formula for the
Cherenkov radiation intensity in a homogeneous medium is obtained. Formula (%
\ref{Omega0}) for the radiation of a charge moving parallel to the axis of
the waveguide can be found, for example, in \cite{Bolo61}.

\begin{figure}[tbph]
\begin{center}
\begin{tabular}{cc}
\epsfig{figure=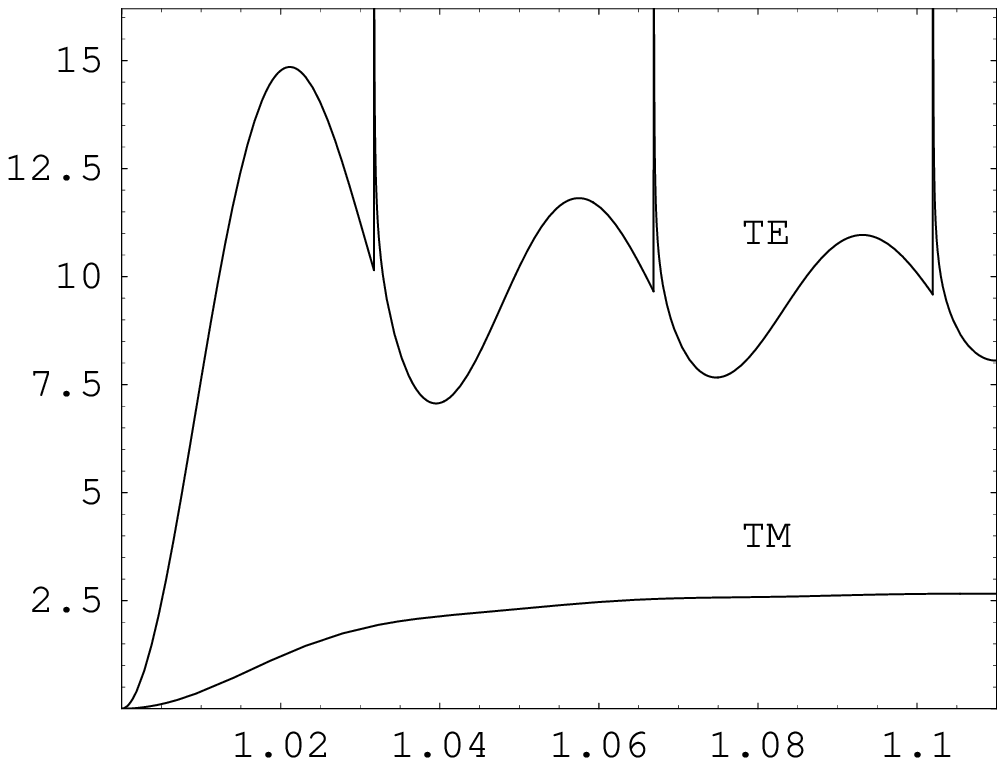,width=7.cm,height=6cm} & \quad %
\epsfig{figure=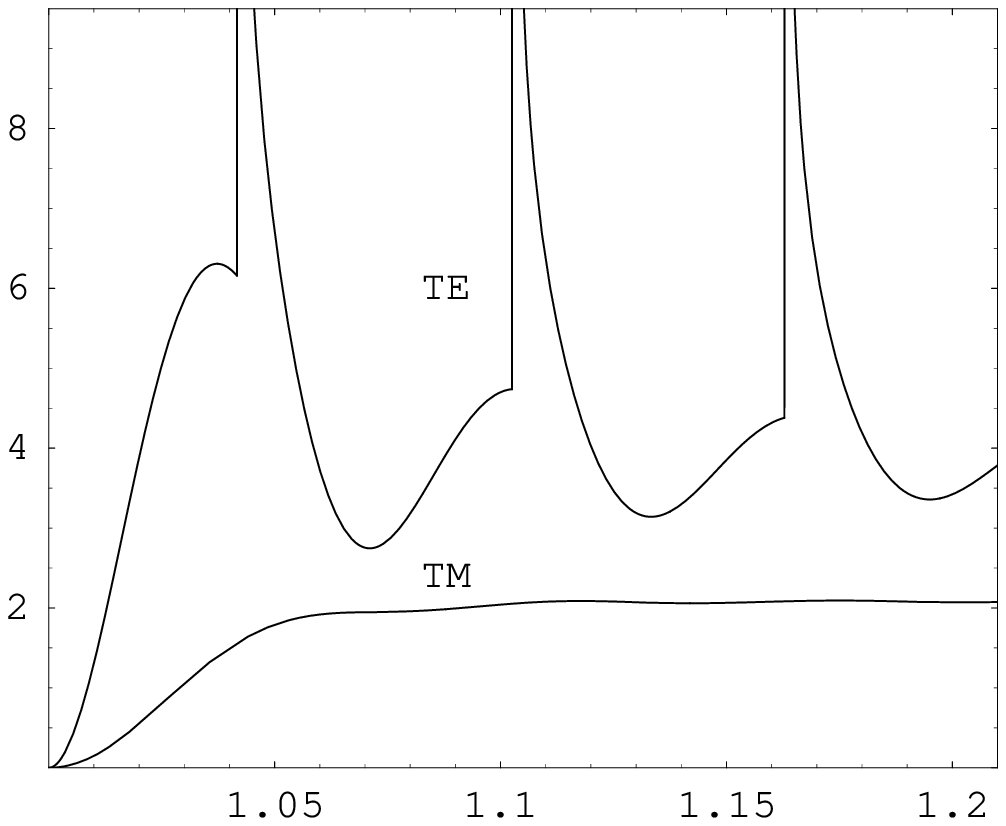,width=7.cm,height=6cm}%
\end{tabular}%
\end{center}
\caption{The number of quanta emitted on the harmonic $m=24$ in the form of
the TE and TM waves per circulating period of the charge, multiplied by $%
\hbar c/q^{2}$, $(\hbar c/q^{2})N_{m}^{(\protect\sigma )}$, versus the ratio
$\protect\rho _{1}/\protect\rho _{0}$ for $\protect\beta _{\parallel }=0.9$
(left panel) and $\protect\beta _{\parallel }=0.7$ (right panel). The
velocity of the transverse motion corresponds to the energy 2 MeV and the
dielectric permittivity is taken $\protect\varepsilon _{0}=3$. }
\label{fig3}
\end{figure}

We have carried out numerical calculations for the number of the radiated
quanta per one period of the particle orbiting,%
\begin{equation}
N_{m}^{(\sigma )}=\sum_{\alpha =\pm }\sum_{n=1}^{n_{\max
}^{(\sigma )}}N_{m,n}^{(\sigma ,\alpha )}=\frac{2\pi }{\hbar
\omega _{0}}\sum_{\alpha =\pm
}\sum_{n=1}^{n_{\max }^{(\sigma )}}\frac{I_{m,n}^{(\sigma ,\alpha )}}{%
|\omega _{m,n}^{(\sigma ,\alpha )}|}.  \label{Nm}
\end{equation}%
As it has been mentioned before, the quantity $N_{m,n}^{(\sigma ,\alpha )}$
does not depend on $\alpha $. In figure \ref{fig3} we have plotted the
dependence of $N_{m}^{(\sigma )}$ on the ratio $\rho _{1}/\rho _{0}$ for $%
m=24$ and $v_{\perp }/c\approx 0.967$ corresponding to the energy 2 MeV
assuming that $\varepsilon _{0}=3$. The graphs are given for $\beta
_{\parallel }=0.9$ (left panel) and $\beta _{\parallel }=0.7$ (right panel).
Note that the location of the peaks in the number of radiated quanta for TE
waves is determined by formula (\ref{peakcond}). In particular, for large
values $n$ by using the asymptotic formula (\ref{jmnas}) we see that the
distance between the neighboring peaks is given by formula $\pi \sqrt{%
1-\beta _{\parallel }^{2}}/(m\beta _{\perp })$ and decreases with increasing
$\beta _{\parallel }$. As we see the main part of the radiated quanta is
emitted in the form of the TE waves. Similar features take place for the
radiation on other values of the harmonic $m$. For the same values of the
parameters, for the number $N_{0m}^{(\sigma )}$ of the radiated quanta in
the homogeneous medium evaluated from formulae (\ref{ImhamTM}), (\ref%
{ImhamTE}), one finds%
\begin{eqnarray}
N_{0m}^{(0)} &\approx &2.683\frac{q^{2}}{\hbar c},\;N_{0m}^{(1)}\approx 9.514%
\frac{q^{2}}{\hbar c},\;\mathrm{for}\;\beta _{\parallel }=0.9,  \label{hamN1}
\\
N_{0m}^{(0)} &\approx &2.082\frac{q^{2}}{\hbar c},\;N_{0m}^{(1)}\approx 4.676%
\frac{q^{2}}{\hbar c},\;\mathrm{for}\;\beta _{\parallel }=0.7.  \label{hamN2}
\end{eqnarray}%
Of course, in this case the result does not depend on $\rho _{1}$.

\begin{figure}[tbph]
\begin{center}
\begin{tabular}{cc}
\epsfig{figure=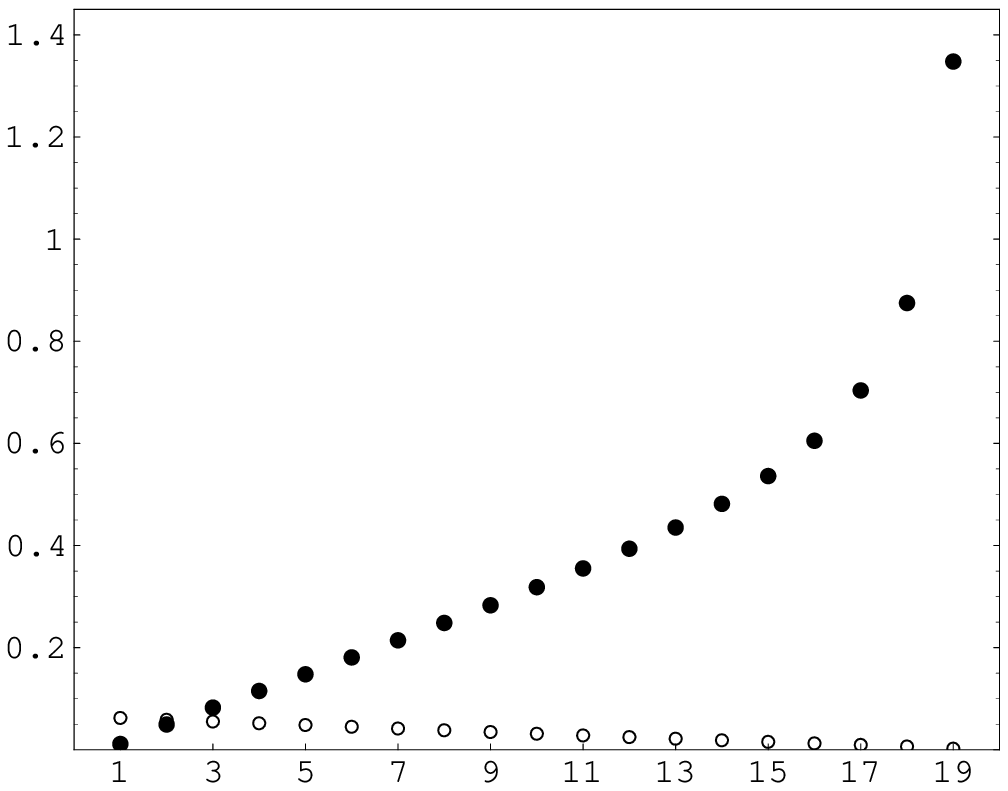,width=7.cm,height=6cm} & \quad %
\epsfig{figure=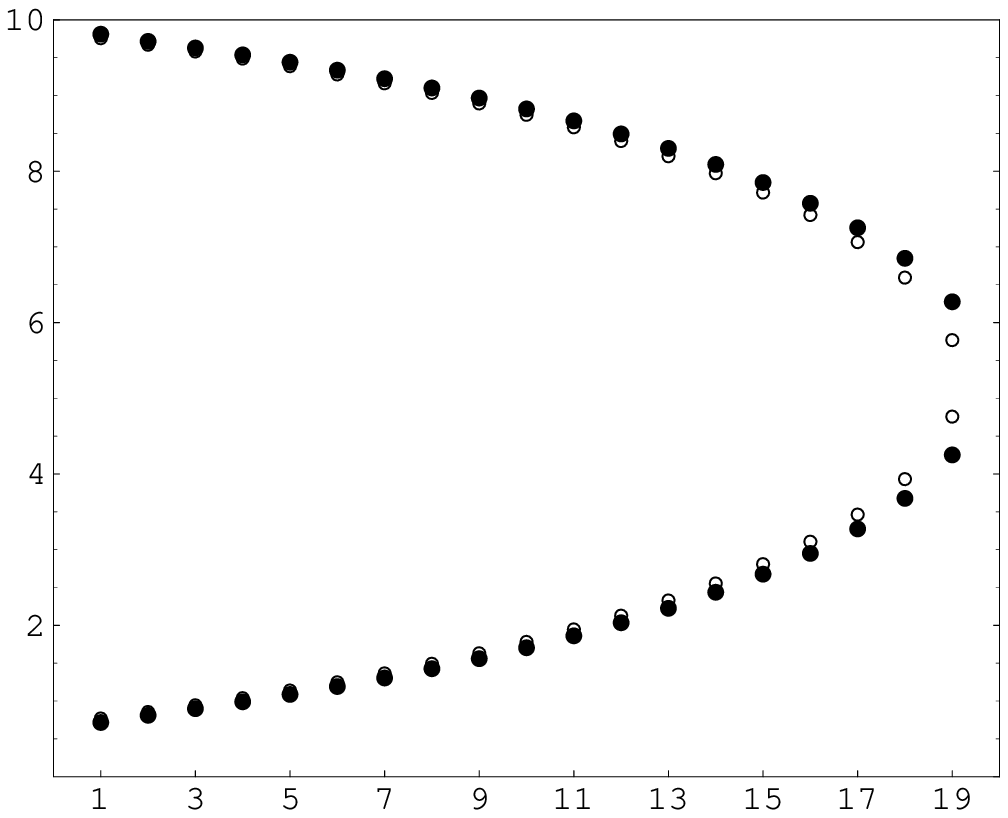,width=7.cm,height=6cm}%
\end{tabular}%
\end{center}
\caption{The left panel presents the number of radiated quanta, $(\hbar
c/q^{2})N_{m,n}^{(\protect\sigma ,\protect\alpha )}$, as a function of $n$
for the harmonic $m=24$ and for $\protect\rho _{1}/\protect\rho _{0}=1.02$, $%
\protect\beta _{\parallel }=0.9$. The values of the other parameters are the
same as those for figure \protect\ref{fig3}. The black points correspond to
the TE waves and the circles correspond to the TM waves. On the right panel
the ratio $\protect\omega _{m,n}^{(\protect\sigma ,\protect\alpha )}/m%
\protect\omega _{0}$ is plotted in dependence of $n$. The upper (lower)
halves of the points correspond to $\protect\alpha =+$ ($\protect\alpha =-$%
). The values of the parameters are the same as for the left panel.}
\label{fig4}
\end{figure}

In addition to the total number of radiated quanta for a given $m$, it is of
interest to consider the corresponding spectral distribution. On the left
panel of figure \ref{fig4} we have plotted the quantity $N_{m,n}^{(\sigma
,\alpha )}$ as a function of $n$ for $\rho _{1}/\rho _{0}=1.02$, $\beta
_{\parallel }=0.9$, for the same values of the parameters corresponding to
figure \ref{fig3}. In this case we have $n_{\max }^{(\sigma )}=19$. On the
right panel of figure \ref{fig4} we have given the corresponding frequencies
for TM and TE waves.

\begin{figure}[tbph]
\begin{center}
\epsfig{figure=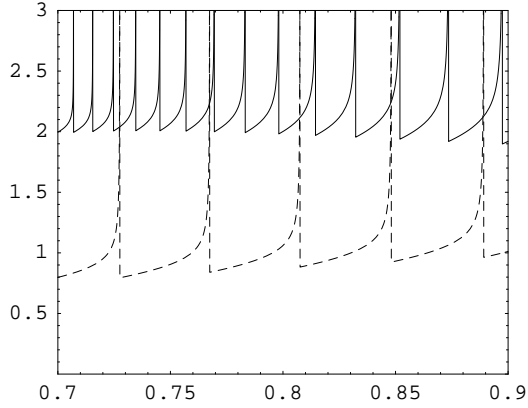,width=7.cm,height=5.5 cm}
\end{center}
\caption{The number of the radiated quanta of the TE modes with $m=1$, $%
(\hbar c/q^2)N_{1}^{(1)}$, as a function of the undulator parameter $K_{%
\mathrm{u}}$ for the undulator period $\protect\lambda _{\mathrm{u}}=3$ cm
and for the radius of the waveguide $\protect\rho _{1}=0.5$ cm. The full
curve corresponds to the radiation from an electron with the energy $%
\mathcal{E}=25$ MeV moving in the waveguide filled by air and the dashed
curve is for the radiation from an electron of energy 100 MeV moving in the
empty waveguide.}
\label{fig5}
\end{figure}

\begin{figure}[tbph]
\begin{center}
\begin{tabular}{cc}
\epsfig{figure=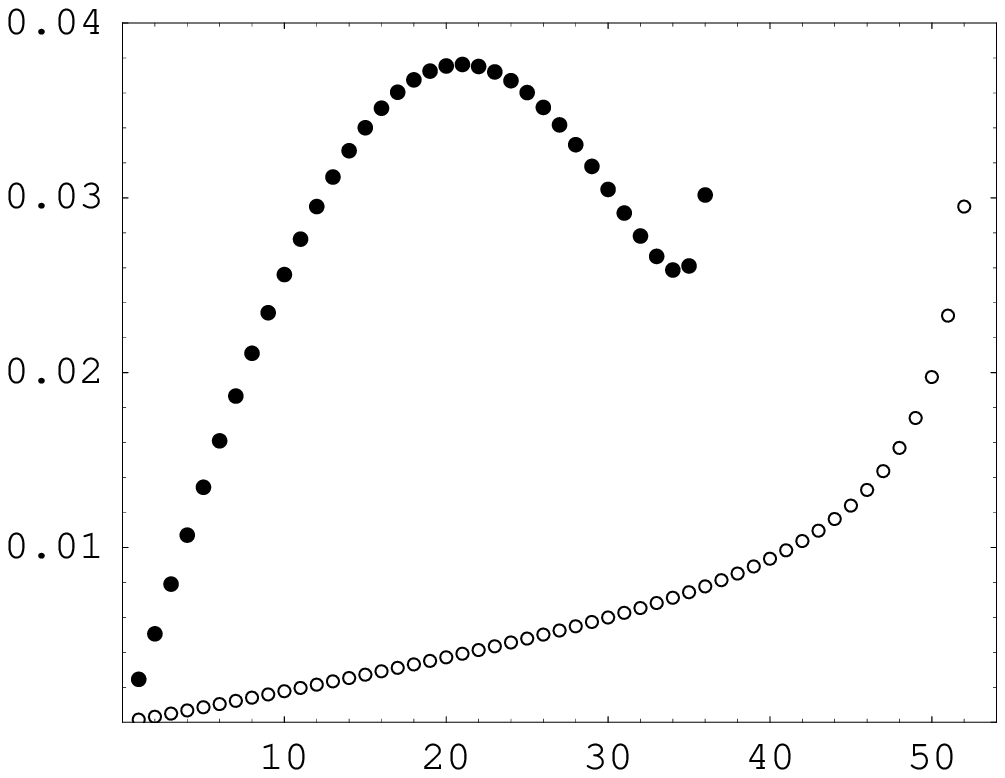,width=7.cm,height=6cm} & \quad %
\epsfig{figure=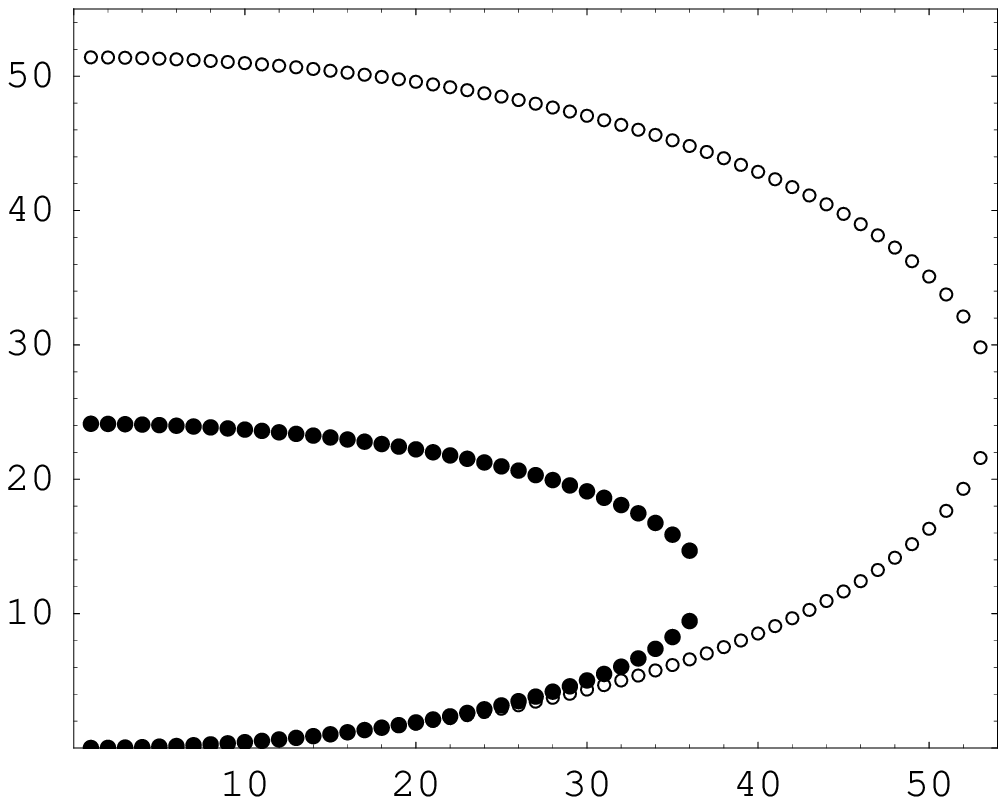,width=6.5cm,height=5.75cm}%
\end{tabular}%
\end{center}
\caption{The left panel presents the number of the radiated quanta, $(\hbar
c/q^2)N_{1,n}^{(1,\protect\alpha )}$, as a function of $n$ for $K_{\mathrm{u}%
}=0.7$. The values of the other parameters are the same as those for figure
\protect\ref{fig5}. On the right panel the corresponding frequencies, $%
10^{-13}\protect\omega _{1,n}^{(1,\protect\alpha )}/2\protect\pi $, are
plotted versus $n$. The black points correspond to the radiation from an
electron with the energy $\mathcal{E}=25$ MeV moving in the waveguide filled
by air and the circles correspond to the radiation from an electron of
energy 100 MeV in the empty waveguide.}
\label{fig6}
\end{figure}

Now we consider an important case of relativistic charge motion in direction
of the waveguide axis when the velocity of orthogonal motion is
non-relativistic, $v_{\perp }\ll c$. This type of motion is realized in
helical undulators. The corresponding magnetic field in Cartesian
coordinates is given by $\mathbf{H}_{\mathrm{u}}=H_{\mathrm{u}}(-\sin (k_{%
\mathrm{u}}z),\cos (k_{\mathrm{u}}z),0)$, where $k_{\mathrm{u}}=2\pi
/\lambda _{\mathrm{u}}$ and $\lambda _{\mathrm{u}}$ is the undulator period
length. The corresponding parameters for the particle orbit are related to
the particle energy $\mathcal{E}$ and to the undulator characteristics by
the formulae%
\begin{equation}
\frac{v_{\perp }}{c}=\frac{K_{\mathrm{u}}}{\gamma },\;\frac{v_{\parallel }}{c%
}=\sqrt{1-\frac{1+K_{\mathrm{u}}^{2}}{\gamma ^{2}}},\;\omega _{0}=k_{\mathrm{%
u}}v_{\parallel },\;\rho _{0}=\frac{K_{\mathrm{u}}c}{\gamma k_{\mathrm{u}%
}v_{\parallel }},  \label{HelUnd}
\end{equation}%
where $\gamma =\mathcal{E}/m_{0}c^{2}$, with $m_{0}$ being the particle
mass. In formulae (\ref{HelUnd}), $K_{\mathrm{u}}=(q/m_{0}c^{2})H_{\mathrm{u}%
}/k_{\mathrm{u}}$ is the so-called undulator parameter. For example, for the
helical undulator of the Stanford free electron laser $H_{\mathrm{u}}=0.23$
T, $\lambda _{\mathrm{u}}=3.3$ cm and the electron energy $\mathcal{E}=43.5$
MeV. For these values of the parameters we have $K_{\mathrm{u}}\approx 0.71$%
. As it is seen from (\ref{ITM}), (\ref{ITE}), in the presence of
the medium the factor $1-v_{\parallel }^{2}/c^{2}$ in the formulae
for the radiation intensity in the empty waveguide is replaced by
the factor $1-\beta _{\parallel }^{2}$. This replacement leads to
important influences on the radiation properties. These influences
are essentially different in the cases $\varepsilon _{0}<1$ and
$\varepsilon _{0}>1$. In the first case, even at very high
energies the factor $1-\beta _{\parallel }^{2}$ tends to finite
limiting value and the radiation does not have the features
typical for the radiation of an ultrarelativistic particle in
vacuum. In contrast to this, when $\varepsilon _{0}>1$, under the
condition $0<1-\beta _{\parallel }^{2}\ll 1$ the properties of the
radiation are similar to those for the radiation in vacuum from an
ultrarelativistic particle even in the case when $v_{\parallel }$
is not too close to $c$. In order to illustrate these features, in
figure \ref{fig5} we have presented the number of the radiated
quanta of the TE modes with $m=1$ as a function of the undulator parameter $%
K_{\mathrm{u}}$ for the undulator period $\lambda _{\mathrm{u}}=3$ cm and
for the radius of the waveguide $\rho _{1}=0.5$ cm. The full curve
corresponds to the radiation from an electron with the energy $\mathcal{E}=25
$ MeV moving in the waveguide filled by air ($\varepsilon _{0}=1.00054$) and
the dashed curve is for the radiation from an electron of energy 100 MeV
moving in the empty waveguide ($\varepsilon _{0}=1$). Note that in the first
case we have $1-\beta _{\parallel }\approx 4.1\times 10^{-5}$ and in the
second case $1-\beta _{\parallel }\approx 1.9\times 10^{-5}$.
The corresponding spectral distributions are presented in figure \ref{fig6}
for the value of the undulator parameter $K_{\mathrm{u}}=0.7$. On the left
panel we have plotted the quantity $N_{1,n}^{(1,\alpha )}$ as a function of $%
n$ and on the right panel the corresponding frequencies are presented. The
black points correspond to the radiation from an electron with the energy $%
\mathcal{E}=25$ MeV moving in the waveguide filled by air and the circles
correspond to the radiation from an electron of energy 100 MeV in the empty
waveguide.

For ultrarelativistic particles the main part of the radiated
energy is in the spectral range $\omega >\omega _{p}$ with $\omega
_{p}$ being the plasma frequency, where the dielectric
permittivity is well approximated by the formula $\varepsilon
_{0}\approx 1-\omega _{p}^{2}/\omega ^{2}$. In this regime the
radiation frequencies are of the order $\omega _{0}\gamma ^{2}$
and for the undulator parameter $\lambda _{u}\sim 1$ cm we have
$\gamma \gtrsim 10^{3}$. In this range the perfect conductor
boundary condition on the waveguide walls are no longer valid and
for frequencies larger than the corresponding plasma frequency the
waveguide becomes transparent. The corresponding radiation
intensity propagating in the exterior region is strongly directed
in the forward direction and is described by the formulae given in
\cite{Saha05}. As regards the frequency range $\omega \ll \omega
_{p}$ the corresponding radiation intensities are described by formulae (\ref%
{ITM}), (\ref{ITE}), where for materials with $\varepsilon _{0}-1\gg \gamma
^{-2}$ we can substitute $1-\beta _{\parallel }^{2}\approx 1-\varepsilon _{0}
$. By taking into account formula (\ref{HelUnd}) for the radius of the
helical orbit, from formulae (\ref{ITM}), (\ref{ITE}) we can see that for $%
j_{m,n}^{(\sigma )}\ll \gamma $ the number of the radiated quanta $%
N_{m,n}^{(\sigma ,\alpha )}$ is suppressed by the factor $\gamma ^{-2}$ for
both TM and TE waves, whereas the corresponding frequencies are practically
independent on the particle energy. For the modes with $j_{m,n}^{(\sigma
)}\gtrsim \gamma $ we have $|\omega _{m,n}^{(\sigma ,\alpha )}|\approx
cj_{m,n}^{(\sigma )}/\sqrt{1-1/\varepsilon _{0}}\rho _{1}$ and $%
N_{m,n}^{(\sigma ,\alpha )}\sim \gamma ^{-2\sigma }$. In this case the
radiation on the TM modes dominates.

In the discussion above we have considered the radiation emitted by a single
particle moving along a prescribed trajectory. From the point of view of
practical applications the generalization of the obtained results for the
case of the radiation from an electron beam is a next important step. In
helical undulators the beam is bunched and the features of the radiation
critically depend on the ratio of the bunch length to the wavelength of the
emitted radiation. If the wavelength is smaller than the bunch length the
particles in the bunch are not phase correlated and the total power radiated
by the bunch is the sum of single particle parts. In the opposite limit,
called the coherent spontaneous radiation regime, the wavelength of the
radiation exceeds the length of the bunch and the particles radiate
coherently. In this case the radiated power is enhanced by the factor of
number of particles in the bunch relative to the incoherent radiation at the
same wavelength. Here it is important to take into account that long
wavelength radiation is suppressed by the waveguide cutoff condition and
this leads to the constraints for the bunch length in order to have coherent
radiation in the waveguide. Once the fields are evaluated, another
interesting question is related to the influence of the radiation field on
the motion of the radiating particles. The interaction of the beam with its
own radiation can induce an additional microbunching with the possibility of
coherent radiation from particles of the same microbunch. These questions
require a separate consideration and we plan to address them in the future
work. A recent discussion of the beam dynamics in undulators can be found in
\cite{Onuk03} (see also \cite{Ng06}).

Another point which deserves a separate investigation is the role
of the other processes of the particle interaction with the medium
(see also the discussion in \cite{Bell06}). In particular, they
include ionization energy losses, particle bremsstrahlung in
media, and the multiple scattering (see, for instance,
\cite{Term72,Akhi96}). The relative role of these processes
depends on the particle energy and characteristics of the medium.
To our best knowledge, the previous investigations in this
direction were mainly concerned with the case of an unbounded
homogeneous medium and the investigation of the effects induced by
the presence of the waveguide requires a separate consideration.
However, some general features can be obtained by using the
corresponding results for a homogeneous medium. First of all, as
the above-mentioned processes are absent in the vacuum, the
radiation discussed in the present paper is the main mechanism of
the energy losses in sufficiently rarefied medium. Next, while the
ionization losses by an electron rise logarithmically with energy
and bremsstrahlung losses arise nearly linearly, the undulator
radiation intensity arises quadratically and, hence, dominates at
sufficiently high energies. For a particle moving in air the
ionization losses dominate those due to the bremsstrahlung for the
energies less than 100 MeV. By using the standard formula, it can
be seen that in the example corresponding to figures \ref{fig5}
and \ref{fig6} the relative energy loss is $\approx 1\%$ per
meter. An interesting possibility to escape ionization losses in
the medium was indicated in \cite{Bolo61} (see also
\cite{Ginz89}). In this paper it was argued that a narrow empty
channel along the particle trajectory in the solid dielectric does
not affect the radiation intensity if the channel radius is less
than the radiation wavelength. From the other side, the maximum
impact parameter for
ionization losses is of the order $b_{\max }\sim \min [(\hbar /m_{0}c)%
\mathcal{E}/I,c/\omega _{p}]$, where $I$ is the mean excitation energy for
the atom of the medium, and for the channel radius larger than $b_{\max }$
ionization losses are suppressed. As in other processes involving multiple
scattering, we expect that this effect will appear in the formula for the
radiation intensity at given frequency in the form of the additional
multiplicative factor and there exists a critical energy of the particle
below which the multiple scattering does not effect the radiation. We can
try to estimate this factor by using the Migdal formula. For the values of
the parameters taken in the example above this factor leads to the decrease
of the radiation intensity by $\approx 0.5\%$.

\section{Conclusion}

\label{sec:Conc}

We have investigated the electromagnetic field generated by a charge moving
along a helical orbit inside a circular waveguide with dielectric filling.
This type of motion is involved in magnetic devices called helical
undulators which are inserted into a straight sector of storage rings. The
helical undulators are used to generate circularly polarized intense
electromagnetic radiation in a relatively narrow bandwidth. The frequency of
radiation is tunable by varying the beam energy and the magnetic field. In
this paper we have seen that the insertion of a waveguide into the helical
undulator provides an additional mechanism for tuning the characteristics of
the emitted radiation by choosing the parameters of the waveguide and
filling medium. The electric and magnetic fields are presented as the sum of
two parts. The first one corresponds to the fields of the charge in the
homogeneous medium and the second one is induced by the presence of the
waveguide. The Fourier components of the latter are given by formulae (\ref%
{Hml1w}), (\ref{Hmz1w}), (\ref{electricb}), (\ref{electricc}). In order to
extract from the total fields the parts corresponding to the radiation, we
have investigated analytic properties of the Fourier components as functions
on $k_{z}$. These components have poles on the eigenmodes of the waveguide.
We have specified the ways by which these poles should be circled in the
integral over $k_{z}$. For the cases $\beta _{\parallel }<1$ and $\beta
_{\parallel }>1$ the corresponding contours are plotted in figures \ref{fig1}
and \ref{fig2}. By using the residue theorem, the radiation fields are
presented as a superposition of the eigenmodes of the waveguide
corresponding to the TM and TE waves. The projection of the wave vector on
the waveguide axis and the corresponding frequency are given by formulae (%
\ref{kmnpm}), (\ref{ommnsigw}). We have derived formulae (\ref{ITM}) and (%
\ref{ITE}) for the radiation intensity emitted in the form of the
TM and TE waves. Limiting cases are considered and features of the
radiation are investigated. In particular, we have seen that the
main part of the radiated quanta is emitted in the form of the TE
waves. Applications of general formulae to helical undulators are
given. In particular, we have demonstrated that in the case of
filled waveguide the radiation with features characteristic for
ultrarelativistic particles in the empty waveguide is obtained for
moderately relativistic particles. The radiation emitted on the
waveguide modes propagates inside the cylinder and the waveguide
serves as a natural collector for the radiation. This eliminates
the necessity for focusing to achieve a high-power spectral
intensity. The geometry considered here is of interest also from
the point of view of generation and transmitting of waves in
waveguides, a subject which is of considerable practical
importance in microwave engineering and optical fiber
communications.

\section*{Acknowledgement}

The authors are grateful to Professor A.R. Mkrtchyan for general
encouragement and to Professor L.Sh. Grigoryan, S.R. Arzumanyan, H.F.
Khachatryan for stimulating discussions. The work has been supported by
Grant No.~1361 from Ministry of Education and Science of the Republic of
Armenia.


\bigskip

\begin{thebibliography}{99}
\bibitem{Soko86} A.A.~Sokolov and I.M.~Ternov, \textit{\ Radiation from
Relativistic Electrons} (ATP, New York, 1986).

\bibitem{Bord99} \textit{Synchrotron Radiation Theory and Its Development},
edited by V.A.~Bordovitsyn (World Scientific, Singapore, 1999).

\bibitem{Hofm04} A. Hofman, \textit{The Physics of Sinchrotron Radiation}
(Cambridge University Press, Cambridge, 2004).

\bibitem{Bala79} V.E. Balakin and A.A. Mikhailichenko, Preprint INP 79-85
(1979).

\bibitem{Alfe74} D.F. Alferov, Yu.A. Bashmakov, and E.G. Bessonov, Sov.
Phys. Tech. Phys. \textbf{18}, 1335 (1974).

\bibitem{Kinc77} B.M. Kincaid, J. Appl. Phys. \textbf{48}, 2684 (1977).

\bibitem{Luch90} P. Luchini and H. Motz, \textit{Undulators and
Free-electron Lasers} (Clarendon, 1990).

\bibitem{Nikitin} M.M. Nikitin and V.Ya. Epp, \textit{Undulator Radiation}
(Energoatomizdat, Moscow, 1988, in Russian).

\bibitem{Ryb79} G.B. Rybicky and A.P. Lightman, \textit{Radiative Processes
in Astrophysics} (J. Wiley, New York, 1979).

\bibitem{Rull98} P. Rullhusen, X. Artru, and P. Dhez, \textit{Novel
Radtiation Sources Using Relativistic Electrons} (World Scientific,
Singapore, 1998).

\bibitem{Tsytovich} V.N. Tsytovich, Westnik MGU \textbf{11}, 27 (1951, in
Russian).

\bibitem{Grigoryan1995} L.Sh. Grigoryan, A.S. Kotanjyan, and A.A. Saharian,
Izv. Akad. Nauk Arm. SSR Fiz. \textbf{30}, 239 (1995) [Sov. J. Contemp.
Phys. \textbf{30}, 1 (1995)].

\bibitem{Grig95b} S.R. Arzumanian, L.Sh. Grigoryan, Kh.V. Kotanjyan, and
A.A. Saharian, Izv. Akad. Nauk Arm. SSR Fiz. \textbf{30}, 106 (1995) [Sov.
J. Contemp. Phys. \textbf{30}, 12 (1995)].

\bibitem{Grigoryan1998} L.Sh. Grigoryan, H.F. Khachatryan, and S.R.
Arzumanyan, Izv. Akad. Nauk Arm. SSR Fiz. \textbf{33}, 267 (1998) [Sov. J.
Contemp. Phys. \textbf{33}, 1 (1998)], cond-mat/0001322.

\bibitem{Kot2000} A.S. Kotanjyan, H.F. Khachatryan , A.V. Petrosyan, and
A.A. Saharian, Izv. Akad. Nauk Arm. SSR Fiz. \textbf{35}, (2000) [Sov. J.
Contemp. Phys. \textbf{35}, 1 (2000)].

\bibitem{Kot2001} A.S. Kotanjyan and A.A. Saharian, Izv. Akad. Nauk Arm. SSR
Fiz. \textbf{36}, 310 (2001) [Sov. J. Contemp. Phys. \textbf{36}, 7 (2001)].

\bibitem{KotNIMB} A.S. Kotanjyan, Nucl. Instrum. Methods \textbf{B201}, 3
(2003).

\bibitem{Saha03} A.A. Saharian and A.S. Kotanjyan, Izv. Akad. Nauk Arm. SSR
Fiz. \textbf{38}, 288 (2003).

\bibitem{Saha04} A.A. Saharian and A.S. Kotanjyan, Nucl. Instrum. Methods
\textbf{B226}, 351 (2004).

\bibitem{Gevo84} L.A. Gevorgian and P.M. Pogosian, Izv. Akad. Nauk Arm. SSR
Fiz. \textbf{19}, 239 (1994).

\bibitem{Soko68} A.A.~Sokolov and I.M.~Ternov, Zs. Phys. \textbf{211}, 1
(1968).

\bibitem{Saha05} A.A. Saharian and A.S. Kotanjyan, J. Phys. A \textbf{38},
4275 (2005).

\bibitem{Saha06} A.A. Saharian, A.S. Kotanjyan, and M.L. Grigoryan, J. Phys.
A \textbf{40}, 1405 (2007).

\bibitem{Bell06} S. Bellucci and V.A. Maisheev, J. Phys.: Condens. Matter
\textbf{18}, S2083 (2006).

\bibitem{Kara77} G.G. Karapetyan, Izv. Akad. Nauk Arm. SSR Fiz. \textbf{12},
186 (1977).

\bibitem{Jackson} J.D. Jackson, \textit{Classical Electrodynamics} (John
Wiley \& Sons, 1998).

\bibitem{Abramovic} \textit{Handbook of Mathematical Functions}, edited by
M. Abramowitz and I.A. Stegun (Dover, New York, 1972).

\bibitem{Kota02} A.S. Kotanjyan and A.A. Saharian, Mod. Phys. Lett. A
\textbf{17}, 1323 (2002).

\bibitem{Bolo61} B.M. Bolotovskii, Sov. Phys. Usp. \textbf{4}, 781 (1961).

\bibitem{Onuk03} \textit{Undulators, Wigglers and Their Applications,}
edited by H. Onuki and P. Elleaume (Taylor \& Francis, London, 2003).

\bibitem{Ng06} K.Y. Ng, \textit{Physics of Intensity Dependent Beam
Instabilities} (World Scientific, Singapore, 2006).

\bibitem{Term72} M.L. Ter-Mikaelian, \textit{High Energy Electromagnetic
Processes in Condensed Media} (Wiley, New York, 1972).

\bibitem{Akhi96} A.I. Akhiezer and N.F. Shul'ga, \textit{High Energy
Electrodynamics in Matter} (Gordon and Breach, Amsterdam, 1996).

\bibitem{Ginz89} V.L. Ginzburg, \textit{Applications of Electrodynamics in
Theoretical Physics and Astrophysics }(Gordon and Breach, New York, 1989).
\end{thebibliography}
\end{document}